\documentclass[
 twocolumn,superscriptaddress,aps,pra,amsmath,nofootinbib,amssymb,floatfix]{revtex4-2}
\usepackage{graphicx}
\usepackage{amsmath}
\usepackage{amssymb}
\usepackage{amsfonts}
\usepackage{mathtools}
\usepackage{color}
\usepackage{dsfont}
\usepackage{float}
\usepackage{soul}
\usepackage{xr-hyper}
\usepackage{hyperref}
\usepackage{tikz}

\usetikzlibrary{positioning,shapes}

\definecolor{lime}{HTML}{A6CE39}
\DeclareRobustCommand{\orcidicon}{%
	\begin{tikzpicture}
	\draw[lime, fill=lime] (0,0) 
	circle [radius=0.16] 
	node[white] {{\fontfamily{qag}\selectfont \tiny ID}};
	\draw[white, fill=white] (-0.0625,0.095) 
	circle [radius=0.007];
	\end{tikzpicture}
	\hspace{-2mm}
}
\foreach \x in {A, ..., Z}{%
	\expandafter\xdef\csname orcid\x\endcsname{\noexpand\href{https://orcid.org/\csname orcidauthor\x\endcsname}{\noexpand\orcidicon}}
}

\begin{document}
\title{Particle creation in left-handed metamaterial transmission lines}
\date{\today}

\author{Alessandro Ferreri \orcidA{}}
\affiliation{Institute for Quantum Computing Analytics (PGI-12), Forschungszentrum J\"ulich, 52425 J\"ulich, Germany}
\author{David Edward Bruschi \orcidC{}}
\affiliation{Institute for Quantum Computing Analytics (PGI-12), Forschungszentrum J\"ulich, 52425 J\"ulich, Germany}
\affiliation{Theoretical Physics, Universit\"at des Saarlandes, 66123 Saarbr\"ucken, Germany}
\author{Frank K. Wilhelm \orcidB{}}
\affiliation{Institute for Quantum Computing Analytics (PGI-12), Forschungszentrum J\"ulich, 52425 J\"ulich, Germany}
\affiliation{Theoretical Physics, Universit\"at des Saarlandes, 66123 Saarbr\"ucken, Germany}

\begin{abstract}
Transmission lines are excellent examples of quantum simulators of quantum fields. By appropriately driving specific circuit elements, these devices can reproduce relativistic and quantum such as particle creation due to the non-adiabatic stimulation of the quantum vacuum.
We investigate particle creation in left-handed transmission lines induced by the modulation of the Josephson energy in superconducting quantum interference devices. Our results show that, as a consequence of the peculiar dispersion relations present in these systems, particle production occurs with much more favorable conditions with respect to the usual right-handed transmission lines.
\end{abstract}

\maketitle

\section{Introduction}
Quantum field theory in flat \cite{schwartz2014quantum} and curved \cite{birrell1984quantum,Ford_2021,martin-martinez_entanglement_2014} spacetime is the best available
mathematical apparatus that underpins our comprehension of relativistic and quantum
many-body phenomena. 
Among its most successful predictions one finds highly energetic processes such as the dynamical Casimir effect (DCE) \cite{dodonov_fifty_2020,law_resonance_1994,dodonov_generation_1996, crocce_resonant_2001}, the Unruh effect \cite{PhysRevD.14.870,RevModPhys.80.787,Fewster_2016,PhysRevA.82.042332} and the Hawking radiation \cite{hawking1974black,PhysRevLett.85.5042,RevModPhys.93.035002}.
Unfortunately, due to the considerable energy scales and the extreme conditions necessary to witness such phenomena, their direct observation  has not been possible to date. Nevertheless, in the last decade the impressive advancements in quantum technologies based on quantum simulation platforms have led to the successful fabrication of devices that well mimic the main features of such highly energetic phenomena \cite{wilson_photon_2010,wilson_observation_2011,steinhauer2016observation, PhysRevA.69.033602, munoz_de_nova_observation_2019,shi_quantum_2023}. 

Quantum simulators are powerful tools for the study of quantum processes whose reproduction and control in laboratory is, in many case, unfeasible \cite{RevModPhys.86.153,you_atomic_2011}.
Among the vast range of possible quantum simulating devices, here we want to focus our attention on one specific class of them, which can well describe the dynamics of quantum fields in non-adiabatic scenarios, namely superconducting circuits based on transmission lines (TLs) \cite{yurke_quantum_1984,blais_cavity_2004, blais_circuit_2021,Zhang2023simulatinggauge,pozar2011microwave,vool_introduction_2017}.
In the frameworks of quantum field theory and cosmology, these platforms find a large number of theoretical and experimental applications in the engineering of spacetime analogues \cite{universe8090452,terrones_towards_2021,shi_quantum_2023}, as well as the simulation of phenomena of particle creation \cite{nation_colloquium_2012, tian_analog_2017,lang_analog_2019}, such as the DCE \cite{lahteenmaki_dynamical_2013,johansson_dynamical_2009, johansson_dynamical_2010} and the Hawking radiation \cite{nation_analogue_2009, tian_analogue_2019, blencowe_analogue_2020}.

In general, the dynamics of quantum scalar fields and particle creation phenomena are always described in an environment characterized by positive dielectric constant and magnetic permeability. However, in the last decades a particular devotion has been dedicated to theoretically and experimentally engineering metamaterials characterized by
left-handed dispersion relations \cite{PhysRevLett.84.4184, PhysRevLett.85.2933, PhysRevLett.90.137401}. In left-handed metamaterials, the group velocity has opposite sign with respect to the Poynting vector, which stems from the fact that both the dielectric constant and the magnetic permeability of the medium are negative \cite{veselago1967electrodynamics}.

In one-dimensional systems, such as TLs, the left-handedness emerges when the dispersion law displays a negative group velocity.
A standard left-handed transmission line (LHTL), as that depicted in Fig.\ref{scheme}a, can be assembled starting from the model of a right-handed transmission line (RHTL) by interchanging capacitors and inductors \cite{kozyrev_nonlinear_2008} (see also Fig.\ref{schemeR}a). 
As a consequence, LTHLs can be seen as the dual of the RHTLs \cite{alizadeh_tunable_2020, egger_multimode_2013}.
Among other uses, LTHLs and hybrid platforms can find applications in circuit QED for the simulations of multimode quantum systems \cite{egger_multimode_2013,messinger_left-handed_2019,wang_mode_2019,indrajeet_coupling_2020, mcbroomcarroll2023entangling}.

We propose two platforms based on left-handed transmission lines, whose dynamics is regulated by a set of superconducting quantum interference devices (SQUIDs) \cite{ovchinnikova_design_2013,shramkova_electrically_2017}. The difference between the two schemes relies on the placement, which leads to two different spectra. 
We will explore the dynamics of such TLs by studying the dynamics of the quantum magnetic flux $\hat\Phi$,
showing that we can stimulate the quantum vacuum of the transmission line and generate pairs of bosonic excitations. Interestingly, we can attribute a different physical interpretation to the particle production depending on the placement of the circuit elements.

The paper is structured as follows: in Sec.\ref{classical} we introduce the two LHTLs and calculate the dispersion relations starting from the linear Lagrangians of the circuits. In Sec. \ref{quantum} we present the protocols to quantize the magnetic flux field along the two TLs. In Sec.\ref{particle} we show our main results: the non-adiabatic modulation of the Josephson energy at the SQUIDs stimulates the creation of bosons along the TL. Finally, in Appendix \ref{appA} we briefly discuss standard RHTL present in literature, while in Appendices \ref{Bog} and \ref{MSA} we present details about calculations, with particular focus on the input-output formalism and the multiscale analysis.

\section{Classical LHTLs}\label{classical}
We introduce three LHTLs based on the presence or absence of capacitors and inductors: we refer to Fig.\ref{scheme}a as \textit{standard scheme}, to  Fig.\ref{scheme}b as \textit{circuit 1}, and to Fig.\ref{scheme}c as \textit{circuit 2}. The difference between the last two platforms relies on the placement of the SQUID: \textit{circuit 1} is characterized by a set of SQUID placed in parallel, whereas in \textit{circuit 2} all SQUIDs are placed in series.
Each circuits consist of $N$ cells, each thereof has length $\Delta x$.

We assume that each SQUID is characterized by the capacitance $C_\textrm{J}$, the Josephson energy $E(t)$, and the phase $\varphi=2\pi\Phi_{\textrm{J}}/\phi_0$, where $\Phi_{\textrm{J}}$ is the magnetic flux at the SQUID, and $\phi_0=\pi\hbar/e$ is the magnetic flux quantum. 
Importantly, we will assume small amplitude of the plasma oscillation
in the SQUID, i.e., $\Phi_\textrm{J}/\phi_0\ll 1$, and that all SQUIDs work in the phase regime $E(t)\gg(2e)^2/2C_\textrm{J}$ \cite{johansson_dynamical_2009,weisl_kerr_2015}, thereby expanding the Josephson energy at the lowest order in $\Phi_\textrm{J}/\phi_0$ \cite{PhysRevB.74.224506}. 
Finally, we make the identification $\Phi_{\textrm{J}}=\Phi$, where $\Phi$ is the magnetic flux on the TL \cite{johansson_dynamical_2010,weisl_kerr_2015}.

The Josephson energy can be externally driven in order to have a time-dependent dispersion relation.
In each platform discussed in this work, we will modulate the Josephson energy via $E(t)=E_0[1+4\eta\sin(\Omega t)]$ around the constant value $E_0=I_\textrm{c} \phi_0$ \cite{ dodonov_fifty_2020}, where $I_\textrm{c}$ is critical current, $\Omega$ is the oscillation frequency and $\eta\ll1$ is dimensionless oscillation amplitude. 


\subsection{Circuit 1: SQUIDs in parallel}
\begin{figure}[h]
	\centering	\includegraphics[width=1\linewidth]{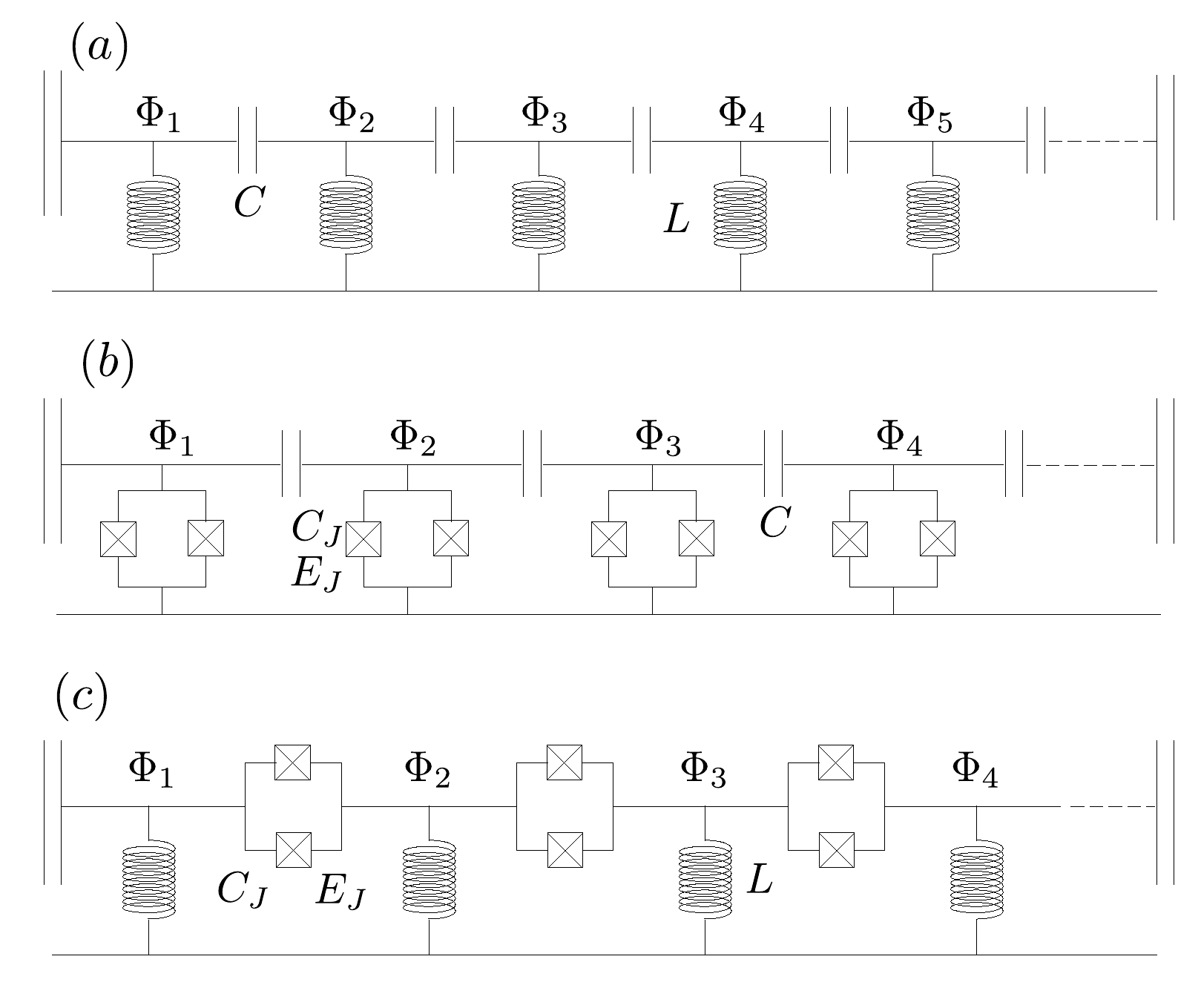}
\caption{Schematic representations of the left-handed transmission line. (a) Standard LHTL. (b) Circuit 1: LHTL with a set of SQUIDs in parallel. (c) Circuit 2: LHTL with a set of SQUIDs in series.}
	\label{scheme}
\end{figure}

As a first case, we want to study the left-handed transmission line pictorially represented in Fig.\ref{scheme}b. At each node we replace the inductors of the standard LHTL shown in Fig.\ref{scheme}a with a SQUID having capacitance and Josephson energy $C_\textrm{J}$ and $E_\textrm{J}=E(t)$ respectively, whereas a capacitor with capacitance $C$ is placed between two nodes.

In the linearized limit of the SQUID \cite{johansson_dynamical_2010,weisl_kerr_2015}, the Lagrangian of the system expressed in terms of the magnetic flux is
\begin{align}
\mathcal{L}=\frac{1}{2}\sum_{n=1}^{N}\left[C(\dot\Phi_n-\dot\Phi_{n+1})^2+C_\textrm{J}\dot\Phi_n^2-\tilde E(t)\Phi_n^2\right],
\label{ls1}
\end{align}
where $\Phi_n$ indicates the magnetic flux field at node $n$, and $\tilde E(t)=(2\pi/\phi_0)^2 E(t)$. 

In order to describe the dynamics of the transmission line we need to solve the equations of motion obtained via the Euler-Lagrange equations $\frac{\textrm{d}}{\textrm{d}t}\partial \mathcal{L}/\partial\dot\Phi_n-\partial \mathcal{L}/\partial\Phi_n=0$. 
These equations admit a set of solutions of the form $e^{i n\Delta x\,k-i\omega t}$. We then compute the dispersion relation of the transmission line, and obtain
\begin{equation}
\omega_j(t)=\sqrt{\frac{\tilde E(t)}{4C\sin^2\left(\frac{k_j\Delta x}{2}\right)+C_{\textrm{J}}}},
\label{ds1}
\end{equation}
where the wave vector, $k_j=\frac{2\pi j}{N\Delta x}$, is bounded within the first Brillouin zone $j=\pm 1,...,\pm N/2$ \cite{egger_multimode_2013}. The ``left-handedness" of the TL clearly emerges from the group velocity defined as $v_\textrm{g}=\partial\omega/\partial k$, which is negative. We note that the infrared limit of the TL is $\omega_\textrm{IR}=\sqrt{\tilde E/(4C+C_{\textrm{J}})}\simeq \sqrt{\tilde E/4C}$, which is reached at the border of the first Brillouin zone, where $k_{N/2}=\pi/\Delta x$, and is valid when $4C\gg C_{\textrm{J}}$ \cite{egger_multimode_2013}.

\subsection{Circuit 2: SQUID in series}
The second scheme we want to analyze is illustrated in Fig.\ref{scheme}c. In this scheme, we replaced each capacitor of the standard LHTL in Fig.\ref{scheme}a  with a SQUID having capacitance $C_{\textrm{J}}$ and Josephson energy $E(t)$. 
The Lagrangian of the TL in the linearized limit of the SQUID is
\begin{align}
\mathcal{L}=&\frac{1}{2}\sum_{n=1}^{N}\left[C(\dot\Phi_{n+1}-\dot\Phi_n)^2-\frac{\Phi_n^2}{L}-\tilde E(t)(\Phi_{n+1}-\Phi_n)^2\right],
\label{ls}
\end{align}
where we made the identification $C\equiv C_\textrm{J}$, and $L$ is the inductance. Assuming as before a solution of the equations of motion of the form $e^{i n\Delta x\,k-i\omega t}$, we find the following dispersion relation
\begin{equation}
\omega_j(t)=\sqrt{\frac{1}{4CL\sin^2\left(\frac{k_j\Delta x}{2}\right)}+\frac{\tilde E(t)}{C}},
\label{ds2}
\end{equation}
where again the wave vector is $k_j=\frac{2\pi j}{N\Delta x}$, with $j=\pm 1,...,\pm N/2$.

Note that this dispersion relation structurally differs from Eq.\eqref{ds1}, and in fact it describes a different left-handed quantum field. 
Indeed, the frequencies in Eq.\eqref{ds2} consist of two parts: the first term inside the square root gives the standard relation between the frequency and wave vector in LHTLs; the second term does not depend on the wave vector and it is identical for each mode. 
In field theory, the presence of the latter is the signature of a massive field \cite{schwartz2014quantum}. Therefore, once we quantize the magnetic field flux, its excitations will behave as massive particles with a left-handed group velocity and time-dependent quadratic mass $M^2(t)\propto \tilde E(t)/C$. Note that, in contrast to the standard (right-handed) dispersion relation of the Klein-Gordon field wherein the massive term of the field becomes negligible in the ultrarelativistic limit ($k\gg M c/\hbar$, with $c$ speed of light), in this LHTL the massive term becomes negligible in long-wavelength modes where $k_j^2\ll 1/[(\Delta x)^2\,CL \tilde E(t)]$.

\section{Quantization procedure}\label{quantum}
In this section we discuss the formalism employed for the quantization of the two LHTLs presented in this work, starting from basic circuit equations. Our goal is to achieve a quantized expression of both the modes of the quantum magnetic flux fields and the Hamiltonian at $t<0$, namely before the beginning of the modulation of the Josephson energy. For this reason, we conveniently omit the time-dependence in the Josephson energy, and write $\tilde E=E_0(2\pi/\phi_0)^2$.

As a first step, we solve the equations of motion of the two TLs. These can be achieved from the two Lagrangians in Eq.\ref{ls1} and Eq.\ref{ls}, respectively for \textit{circuit 1} and \textit{circuit 2}, by means of the Euler-Lagrange equations. 
At any node $1<n<N$, the equations of motion become
\begin{align}
\ddot\Phi_{n+1}+\ddot\Phi_{n-1}-2\ddot\Phi_n-\frac{C_\textrm{J}}{C}\ddot\Phi_n=\frac{\tilde E}{C}\Phi_n,\nonumber\\
\ddot\Phi_{n+1}+\ddot\Phi_{n-1}-2\ddot\Phi_n=\frac{\Phi_n}{CL}+\frac{\tilde E}{C}(2\Phi_n-\Phi_{n+1}-\Phi_{n-1}),
\label{eom2}
\end{align}
for \textit{circuit 1} and \textit{circuit 2} respectively. In the second line of Eq.\eqref{eom2} we made the identification $C_\textrm{J}\equiv C$.
Assuming plane wave solution of the form
$e^{i n\Delta x\,k-i\omega t}$, the magnetic flux is described by the expression
\begin{align}
\Phi(n,t)=\sum_{\lvert j\rvert=1}^{N/2}\left[\phi_j(n,t) a_{j}+\phi^*_j(n,t) a_{j}^*\right],
\label{field}
\end{align}
where the modes are defined by $
\phi_j(n,t\le 0)=\sqrt{\frac{\hbar}{2C N\omega_{0j}}}e^{i(k_j n\Delta x-\omega_{0j} t)}$, and $\omega_{0j}\equiv\omega_j(t\le 0)$ are the mode frequencies (these will correspond to either Eq.\ref{ds1} or Eq.\ref{ds2} depending on the considered scheme). Note that, since the minimum distinguishable wavelength is $\lambda_{\textrm{min}}=2\Delta x$ \cite{egger_multimode_2013}, the sum over all modes runs up to $N/2$. 
The modes are normalized via the relation
{\small
\begin{align}
-\frac{iC}{\hbar}\sum_{n=1}^N\left[\phi_i(n,t)\frac{\partial\phi_j^*(n,t)}{\partial t}-\frac{\partial\phi_i^*(n,t)}{\partial t}\phi_j(n,t)\right]=\delta_{ij},
\label{norm}
\end{align}
}
where we made use of the representation of the Kronecker delta 
$\delta_{lh}=\frac{1}{N}\sum_{n=1}^N e^{2\pi i n(l-h)/N}$.

The Hamiltonian of the transmission lines is achieved from the Lagrangian by means of the Legendre transformation, $\mathcal{H}=\sum_n [P_{n}\dot \Phi_n-\mathcal{L}]$, with conjugated momenta
\begin{align}
P_{n}=&\frac{\partial \mathcal{L}}{\partial\dot \Phi_{n}}=C\left(2\dot\Phi_{n}-\dot\Phi_{{n}+1}-\dot\Phi_{{n}-1}\right)+C_\textrm{J}\dot \Phi_{n},\nonumber\\
P_{n}=&\frac{\partial \mathcal{L}}{\partial\dot \Phi_{n}}=C\left(2\dot\Phi_{n}-\dot\Phi_{{n}+1}-\dot\Phi_{{n}-1}\right),
\label{momentum}
\end{align}
for \textit{circuit 1} and \textit{circuit 2} respectively.

We take advantage of both the mode expansion in Eq.\eqref{field} and the definition of the conjugate momentum in Eq.\eqref{momentum} to perform the discrete Fourier transform of both the field and the conjugate momentum
thereby obtaining the classical amplitude of the field in terms of $\Phi_{n}$ and $P_{n}$. This reads
{\small
\begin{align}
a_h=\zeta_h\sum_n^N e^{-i(k_h n\Delta x-\omega_h t)}\left[\Phi(n,t)+\frac{i\chi_h^{-1}}{\omega_h}P(n,t)\right].
\label{aPP}
\end{align}
}
with $\zeta_h=\sqrt{\frac{\omega_h C}{2\hbar N}}$. The parameter $\chi_h$ strictly depends on the scheme we are considering. In particular, it has the expression 
\begin{equation}
   \chi_h= \left\{ 
\begin{array}{ll}
4\sin^2\left(\frac{k_h\Delta x}{2}\right)+\frac{C_\textrm{J}}{C} &\quad\textrm{for \textit{circuit 1}}, \nonumber\\    
4\sin^2\left(\frac{k_h\Delta x}{2}\right) &\quad\textrm{for \textit{circuit 2}}.
\end{array}\right.
\end{equation}

The quantization of field and the canonical momentum is accomplished by imposing the equal time commutators $[\hat\Phi(n,t),\hat P(m,t)]=i\hbar\delta_{nm}$ and $[\hat \Phi(n,t),\hat \Phi(m,t)]=[\hat P(n,t),\hat P(m,t)]=0$. We then use these commutators, together with Eq.\eqref{aPP} and the discrete representation of the Kronecker delta provided before, to obtain the commutation relation of the quantized amplitude, promoted to annihilation and creation operators
\begin{align}
\bigl[\hat a_j,\hat a_h^\dag\bigr]=\chi_j^{-1}\delta_{jh}.
\label{commaad}
\end{align}
The presence of the factor $\chi_j$ in the denominator is key, as can be seen below.

We can now express the Hamiltonian $\mathcal{\hat H}$ in terms of the ladder operators. By substituting the mode decomposition in Eq.\eqref{field}, exploiting the normalization condition in Eq.\eqref{norm}, and taking advantage of the commutation rule in Eq.\eqref{commaad}, the Hamiltonian reduces to
\begin{align}
\mathcal{\hat H}=&\hbar\sum_{\lvert j\rvert=1}^{N/2}\omega_{0j}\left[\chi_{j}\hat a_j^\dag\hat a_j+\frac{1}{2}\right].
\label{HH}
\end{align}
Note that the commutation relation in Eq.\eqref{commaad} preserves the correspondence principle, as the Heisenberg equation for the annihilation operator reads
\begin{align}
\frac{d\hat a_j}{dt}=\frac{i}{\hbar}[\mathcal{\hat H},\hat a_j]=-i\omega_j\hat a_j.
\end{align}

The form of both the modes at $t<0$ and the Hamiltonian in Eq.\eqref{HH} is valid for both LHTLs under consideration. The difference between the two schemes relies only on the dispersion relations and the Hamiltonian eigenenergies. In particular, the explicit form of the time-dependent eigenenergies $\epsilon^{(\ell)}_j(t)$ is 
{\small
\begin{align}
\epsilon^{(1)}_j(t)=&\hbar\sqrt{\frac{\chi_j\tilde E(t)}{C}},\label{e1}\\
\epsilon^{(2)}_j(t)=&\hbar\sqrt{\frac{\chi_j\tilde E(t)}{C}}\sqrt{\chi_j+\frac{1}{L \tilde E(t)}}.\label{e2}
\end{align}
}
for \textit{circuit 1} and \textit{circuit 2}, respectively.

\begin{figure}[t!]
	\centering
	\includegraphics[width=1\linewidth]{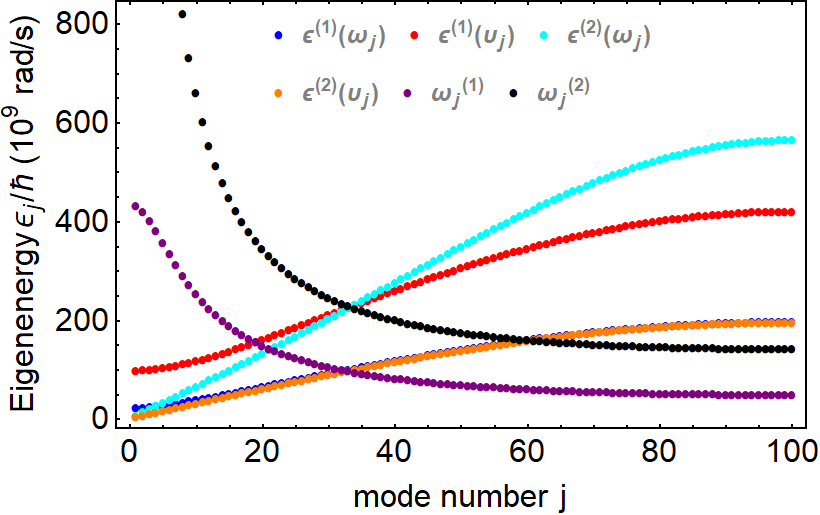}
	\caption{Hamiltonian eigenenergies of the LHTLs and the RHTLs, and dispersion relations of the LHTLs, in circuit 1 and 2. Plots refer to the eigenenergies of the LHTL circuit 1 (blue), and circuit 2 (cyan), eigenenergies (frequencies) of the RHTL circuit 1 (red), and circuit 2 (orange), frequencies of the LHTL circuit 1 (black), and circuit 2 (purple). 
 Indices (1) and (2) label the circuit scheme the plot refers to.
 Parameters are: $C=0.4$ pF, $L=60$ pH, $C_\textrm{J}=0.02$ pF, $I_\textrm{c}=1.25$ $\mu$A and $N=200$.}
 \label{beforetau}
\end{figure}

We notice that, due to this discrepancy between eigenenergies and frequencies in LHTLs, the eigenenergies of LHTLs and RHTLs behave in a similar manner: the lower the wave vectors, the lower the eigenenergies. This is evident in Fig.\ref{beforetau}, where we plot the eigenenergies $\epsilon_j$ of the LHTLs in Eq.\ref{e1} (blue), and Eq.\ref{e2} (cyan), the dispersion relation in Eq.\ref{ds3} (red) and Eq.\ref{ds4} (orange) of the RHTLs illustrated respectively in Fig.\ref{schemeR}b and Fig.\ref{schemeR}c, together with the dispersion relations in Eq.\eqref{ds1} (purple) and Eq.\eqref{ds2} (black). We remind that, unlike LHTLs, the eigenenergies of the Hamiltonian in RHTLs coincide with the mode frequencies, $\epsilon_j(\upsilon_{h})\equiv\upsilon_{h}$.

\section{Particle creation}\label{particle} 
The time-dependent Josephson energy acts as an external drive in the dynamics of the transmission lines. When the modulation frequency $\Omega$ is of the same order of magnitude of the TL mode frequencies, phenomena of particle creation can occur due to the presence of a resonance  \cite{birrell1984quantum}. To describe such phenomena, we need to solve the equation of motion with respect to the magnetic flux field modes or, in other words, we need to find the relation between the annihilation and creation operators of the magnetic flux modes before the beginning of the dynamics (input operators) and the ladder operators obtained once the modulation of the Josephson energy ceases (output operators). Since the Hamiltonian describes linear dynamics for small modulations of the Josephson energy, the input and output ladder operators are related by \textit{Bogoliubov transformations} \cite{birrell1984quantum,Ford_2021,martin-martinez_entanglement_2014}, which have the generic form
\begin{align}\label{bt}
\hat a_{i}^{\textrm{out}}= \sum_{\lvert l\rvert=1}^{N/2}\left( \alpha_{ji}\,\hat a_j^{\textrm{in}}+\beta_{ji}^*\,\hat a_j^{\dag \textrm{in}}\right),
\end{align}
where the explicit expression of the coefficients $\alpha_{ij}$ and $\beta_{ij}$ depend on the specific TL.
A more detailed description of the input-out formalism, as well as the solution of the equations of motion achieved by means of multiscale analysis \cite{bender1999advanced, crocce_resonant_2001}, is reported in Appendix \ref{Bog} and \ref{MSA}.

The phenomenon of particle creation is strongly connected to the coefficients $\beta_{ij}$ in Eq.\eqref{bt}. Indeed, when these coefficients are non-zero, the initial vacuum state of the quantum field does not correspond to the quantum vacuum at the end of the dynamics, and the output operators in Eq.\eqref{bt} do not act as annihilation operators of the initial quantum vacuum \cite{birrell1984quantum}.

In the TLs under consideration, when the modulation frequency of the Josephson energy is exactly twice the frequency of one of the TL modes, $\Omega=2\omega_h$, the coefficient $\beta_{hh}^*$ in Eq.\eqref{bt} does not vanish, and the operator $\hat a_{h}^{\textrm{out}}$ acts on the input state as a squeezed annihilation operator. 
If the system is initially prepared in the vacuum state,
$\lvert 0\rangle=\lvert 0_1, 0_2,...0_{N/2}\rangle$, we can estimate the average number of particles in the resonant mode $\omega_h$ created during the squeezing process as the expectation value of the output number operator $ \hat N_h=(\hat a_h^{\textrm{out}})^\dag \hat a_h^{\textrm{out}}$. Therefore, the output number of particles $\langle\hat N_h(\tau)\rangle$ and the average energy $\langle\mathcal{\hat H}(\tau)\rangle$ at time $\tau=\eta t$ are respectively given by $N_h(\tau)=\sum_j\lvert\beta_{jh}(\tau)\rvert^2=\lvert\beta_{hh}(\tau)\rvert^2$ and $E_h(\tau)=\epsilon_h N_h(\tau)$.

Before studying these quantities numerically, we first notice that the analytical expression for the output photon number $N_h(\tau)$ does not depend on the left- or right-handedness of the TL, but it only depends on the massive or massless character of the Josephson energy in the dispersion relation (see Appendix \ref{MSA}). In particular, the number of particles created is given by 
\begin{subequations}
\begin{equation}
N_h(\tau)=\sinh^2\left(\kappa_{0h}\tau\right),
\label{nmassless}
\end{equation}
\begin{equation}
N_h(\tau)=\sinh^2\left(\frac{\tilde E_0\tau}{C\kappa_{0h}}\right)
\label{nmassive}
\end{equation}
\end{subequations}
for the massless and massive case respectively, with $\kappa=\omega,\upsilon$. 
We notice that, although Eq.\eqref{nmassless} is formally identical to the number of photons created via DCE in a cavity confining a quantum scalar field \cite{dodonov_fifty_2020}, the wave vector of the TLs is not time dependent, therefore the Josephson energy does not simulate the modulation of the cavity length \cite{law_interaction_1995,PhysRevA.106.033502}.  We can provide a more accurate physical interpretation by analyzing the modes in the proximity of the infrared limit, $\omega_j\simeq \omega_\textrm{IR}$. Indeed, in this limit the TL can simulate scenarios wherein the time dependence of the Josephson energy mimics the modulation of the speed of light \cite{lang_analog_2019,sabin_one-dimensional_2018}.

The result in Eq.\eqref{nmassive} can be interpreted through the lens of quantum field theory as the creation of particles due to the modulation of the mass of the quantum field. In right-handed quantum fields, the time modulation of the massive term can be associated to the time-dependence of the metric describing the structure of the spacetime \cite{birrell1984quantum,nation_colloquium_2012, tian_analog_2017}. In case of the LTHL in \textit{circuit 2}, Eq.\eqref{ls} does not correspond to the Lagrangian of a Klein-Gordon scalar field, and the investigation of gravitational effects on  quantum fields with left-handed dispersion relation would require a more comprehensive analysis that is beyond the scope of this work.

\begin{figure}[ht!]
	\centering
	\includegraphics[width=1\linewidth]{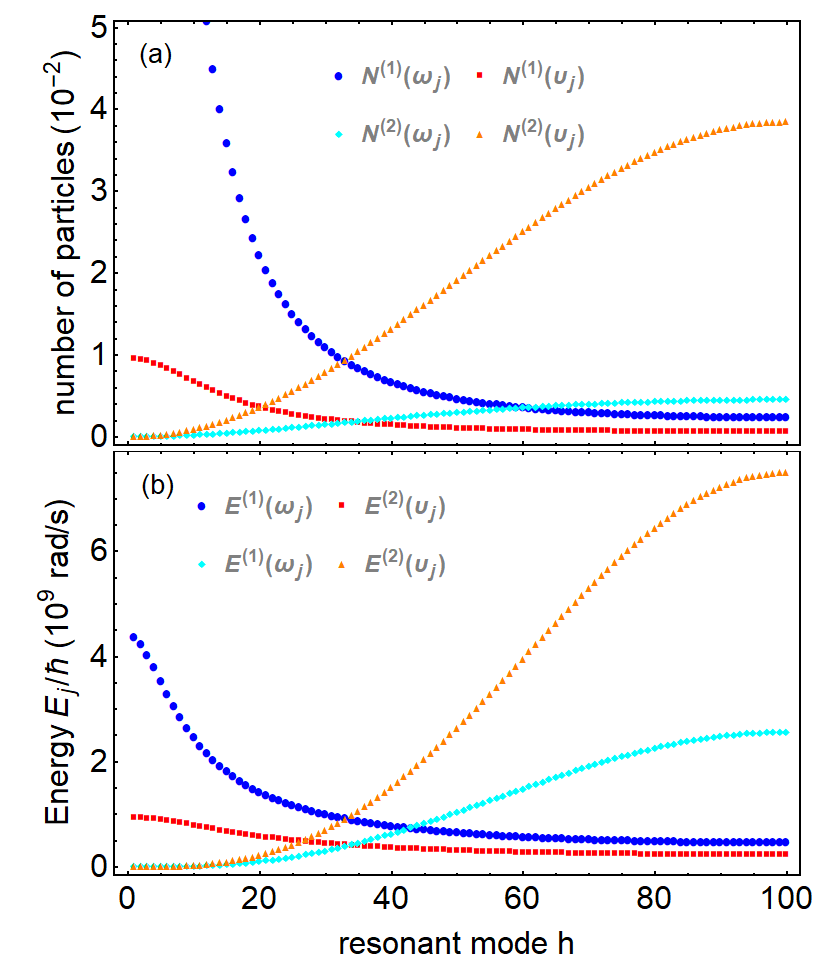}
	\caption{(a) Expected population and (b) energy at time $\tau$ of the mode $j$ by activating the resonance condition $\Omega=2\omega_j$ in the LHTL and $\Omega=2\upsilon_j$ in the RHTL. Plots refer to LHTL circuit 1 (blue), LHTL circuit 2 (cyan), RHTL circuit 1 (red), RHTL circuit 2 (orange).
 Indices (1) and (2) label the circuit scheme the plot refers to.
 Parameters are: $C=0.4$ pF, $L=60$ pH, $C_\textrm{J}=0.02$ pF, $I_\textrm{c}=1.25$ $\mu$A, $N=200$, and $\tau=1$ps.}
 \label{aftertau}
\end{figure}

We now want to study the behavior of $N_h(\tau)$ and $E_h(\tau)$ in the two LTHLs depicted in Fig.\ref{scheme}b and Fig.\ref{scheme}c, as well as in the two RHTLs in Fig.\ref{schemeR}b and Fig.\ref{schemeR}c.
Results of our investigation are plotted in Fig.\ref{aftertau}. In these graphs, each value represents the expected particle number and the energy of the mode $h$ at time $\tau$, assuming that this mode is resonant with the Josephson energy of the SQUID via the resonance condition $\Omega=2\omega_{h}$ (or $\Omega=2\upsilon_{h}$ in the case of RHTLs).

The graph in Fig.\ref{aftertau}a shows the different trend of the particle creation in LHTLs and RHTLs. We observe that, in case of massless excitations, the resonant mode $\upsilon_h$ of the RHTL emits on average more particles at higher mode numbers (orange dots), while in LHTL the number of massless particles in the resonant mode $\omega_h$ drastically increases at lower mode numbers (blue dots). This is due to the fact that the function $N_h(\tau)$ in Eq.\eqref{nmassless} increases with the mode frequency (and not with the mode eigenenergy). The same graph shows that the creation of highly energetic massive particles is higher in LHTLs (cyan dots) than in RHTLs (red dots). 
These results have remarkable consequences for the simulation of particle creation processes.
Indeed this means that, in LHTLs, the creation of massless particles with low energetic modes, as well as the creation of massive particles with highly energetic modes, are both strongly favorable. We notice that the number of massless particles with long wavelengths created via resonant processes is, in contrast, much more difficult to achieve in circuit QED when using standard RHTLs, or in cavity QED via DCE \cite{dodonov_fifty_2020}.

The peculiar functional behaviour of the boson number plotted in Fig.\ref{aftertau}a has also important consequence on the energy released by the system. Although the eigenvalues of the Hamiltonian in both LTHLs and RHTLs displayed strong similarities (see Fig.\ref{beforetau}), Fig.\ref{aftertau}b shows that, at lower mode numbers, the left-handed \textit{circuit 1} releases a higher output energy with respect to the other platforms. In particular, this circuit  well converts the Josephson energy into quanta of the magnetic flux, despite the small eigenenergy of these excitations. Consequently, we can exploit this platform to extract the energy stored in the Josephson junctions of the SQUIDs with higher efficiency.

\section{Conclusion}\label{conclusion}
This work aimed at quantizing left-handed metamaterial transmission lines and studying their quantum dynamics.
We used quantum field theory to quantize the magnetic flux field in LHTLs, providing the commutation rules for the ladder operators and demonstrating the presence of a discrepancy between frequencies and eigenenergies. In this framework we then investigated the creation of particles due to resonant modulation of the Josephson energies, and compared the number of particles generated in LHTLs and RHTLs respectively.
Our results demonstrate not only that particle production in RHTLs and LTHLs show strong mathematical similarities, but also that, due to the peculiar dispersion relation in LHTLs, this phenomenon is drastically amplified in LHTLs for lower energetic modes. The amplification of particle creation at lower energies allows for an easier experimental accessibility to the phenomenon, which can in turn provide several concrete advantages.
For these reasons, we believe that these platforms could find interesting applications in novel quantum technologies, such as sensing and amplification of low-frequency signals. This work also paves the way to future investigations of quantum field simulators based on left-handed metamaterial transmission lines.

\section{Acknowledgments}
A.F., D.E.B and F.K.W. acknowledge support from the joint project No. 13N15685 ``German Quantum Computer based on Superconducting Qubits (GeQCoS)'' sponsored by the German Federal Ministry of Education and Research (BMBF) under the \href{https://www.quantentechnologien.de/fileadmin/public/Redaktion/Dokumente/PDF/Publikationen/Federal-Government-Framework-Programme-Quantum-technologies-2018-bf-C1.pdf}{framework programme
``Quantum technologies -- from basic research to the market''}.
D.E.B. also acknowledges support from the German Federal Ministry of Education and Research via the \href{https://www.quantentechnologien.de/fileadmin/public/Redaktion/Dokumente/PDF/Publikationen/Federal-Government-Framework-Programme-Quantum-technologies-2018-bf-C1.pdf}{framework programme
``Quantum technologies -- from basic research to the market''} under contract number 13N16210 ``SPINNING''.

\bibliographystyle{apsrev4-2}
\bibliography{ref}
\appendix
\onecolumngrid

\section{Right-handed transmission lines}\label{appA}
\begin{figure}[ht]
	\centering	\includegraphics[width=0.7\linewidth]{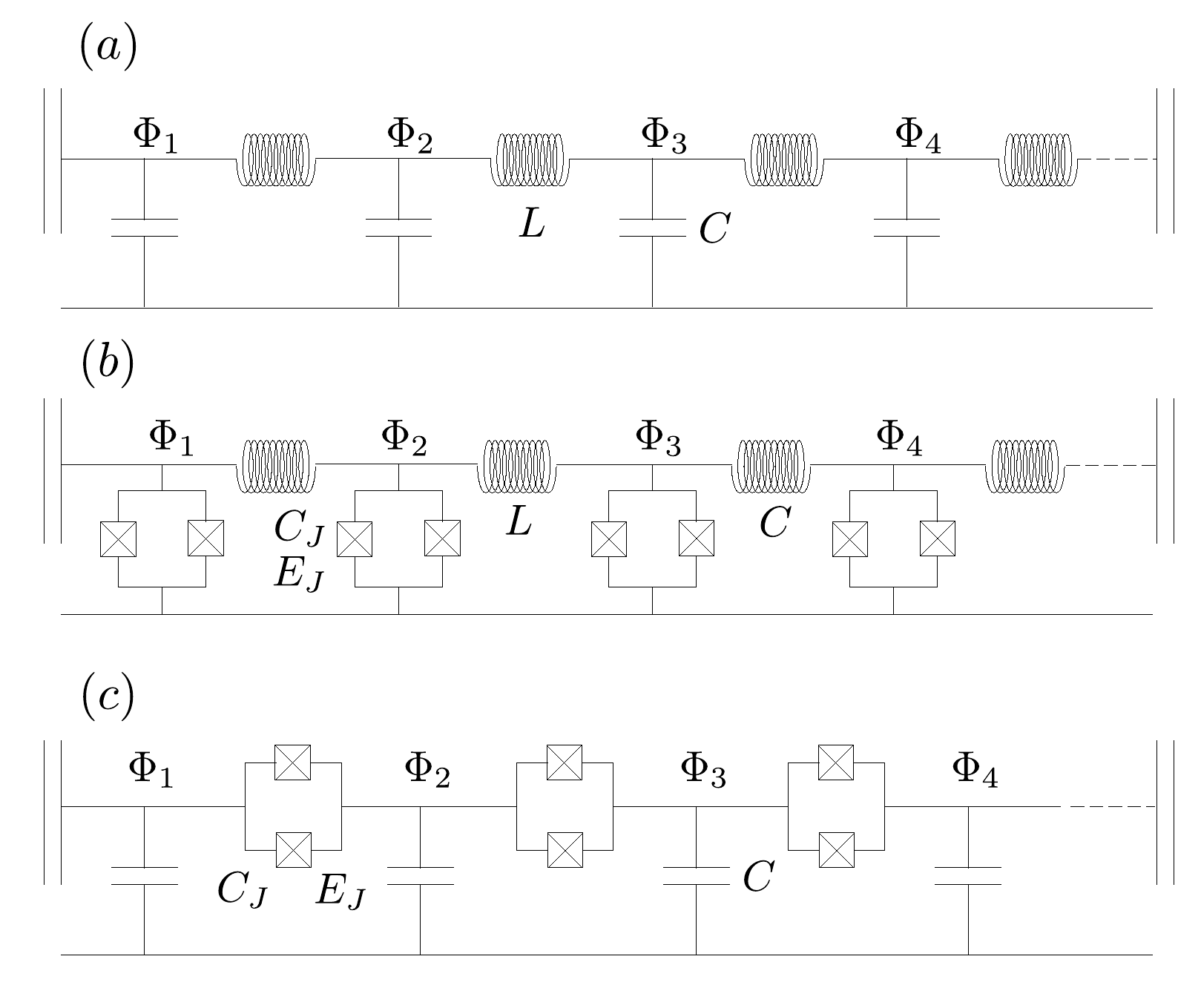}
\caption{Schematic representations of the right-handed transmission line. (a) Standard RHTL. (b) Circuit 1: RHTL with a set of SQUIDs in parallel. (c) Circuit 2: RHTL with a set of SQUIDs in series.}
	\label{schemeR}
\end{figure}


In this section, we want to make a comparison between our two LHTLs and the equivalent RHTLs illustrated in Fig.\ref{schemeR}b and Fig.\ref{schemeR}c. 
The first scheme we want to consider, which we will name \textit{right-handed circuit 1}, is represented in Fig.\ref{schemeR}b, and it is achieved from the standard RHTL in Fig.\ref{schemeR}a by replacing all capacitors with SQUIDs. A similar circuit was already proposed as analogue of a scalar field interacting with the the gravitational field \cite{tian_analog_2017}. The dispersion relation of such RHTL is
\begin{align}
\upsilon_j(t)=\sqrt{\frac{4\sin^2\left(\frac{p_j\Delta x}{2}\right)}{L C}+\frac{\tilde E(t)}{C}},
\label{ds3}
\end{align}
with $C\equiv C_{\textrm{J}}$. In this work, $\upsilon_j$ and $p_j$ always indicate the frequency and the wave vector in the RHTL, respectively.
In the framework of quantum field theory, this expression suggests that the second term inside the square root is related to the presence of a massive term with mass parameter $M^2(t)\propto\tilde E(t)/C$, see \cite{tian_analogue_2019}. 

The second RHTL, which we will refer to as \textit{right-handed circuit 2}, is illustrated in Fig.\ref{schemeR}c and is implemented from the scheme in Fig.\ref{schemeR}a by replacing all inductors with a set of SQUIDs \cite{masluk_microwave_2012}. A similar scheme was realized for the study of nonlinear effects in the bare frequencies of the Hamiltonian (Kerr effect) \cite{weisl_kerr_2015}. The dispersion relation in this TL reads
\begin{align}
\upsilon_j(t)=&2\sin\left(\frac{p_j\Delta x}{2}\right)\sqrt{\frac{\tilde E(t)}{C+4C_\textrm{J}\sin^2\left(\frac{p_j\Delta x}{2}\right)}}\simeq 2\sin\left(\frac{p_j\Delta x}{2}\right)\sqrt{\frac{\tilde E(t)}{C}},
\label{ds4}
\end{align}
where we assumed $4C_\textrm{J}\ll C$.

The Hamiltonian of RHTLs under consideration takes the same form for both schemes in Fig.\ref{schemeR}b and Fig.\ref{schemeR}c, namely
\begin{align}
\mathcal{\hat H}_\textrm{R}=&\hbar\sum_{\lvert j\rvert=1}^{N/2} \upsilon_j\hat b_j^\dag\hat b_j.
\label{HR}
\end{align}
A direct comparison between the dispersion relations in LHTLs (see Eq.\eqref{ds1} and \eqref{ds2} in the main text), and in RHTLs (see Eqs. \eqref{ds3} and \eqref{ds4}) shows interesting analogies. For instance, the Josephson energy can play the role of a massless or a massive term in both types of TLs. Finally, the fact that the dispersion relation of the \textit{left-handed circuit 1} shares the same form of the dispersion relation of the \textit{right-handed circuit 2}, and analogously for the other two schemes, is a further signature of the duality between left-handed and right-handed transmission lines. 

\section{Dynamics and Bogoliubov transformation}\label{Bog}
To study the dynamics at $t>0$ we will use the procedure employed in \cite{crocce_resonant_2001, bender1999advanced}. Arguments presented in this section are valid for both LHTLs considered in this work. 
When the modulation of the Josephson energy starts, each mode can be written as:
\begin{align}
\phi_j(n,t>0)=\sum_{\lvert l\rvert=1}^{N/2}\sqrt{\frac{\hbar}{C N}}Q_{jl}(t)e^{i k_l n\Delta x},
\label{p>0}
\end{align}
where $Q_{jl}(t)$ are solution of the equations of motion with continuity conditions $Q_{jl}(0)=\delta_{jl}/\sqrt{2\omega_{0j}}$ and $\dot Q_{jl}(0)=-i\delta_{jl}\sqrt{2\omega_{0j}}$, with $\omega_{0j}\equiv\omega_j(t\le 0)$.

The explicit expression for $Q_{jl}(t)$ are achieved by solving the equations of motion. The substitution of Eq.\eqref{p>0} into the equations of motion yields:
\begin{align}
 \sum_{\lvert l\rvert=1}^{N/2}\left[\frac{\ddot Q_{jl}(t)}{\omega_j^2(t)}+Q_{jl}(t)\right]e^{i n\Delta x k_l}&=0.
\end{align}
Multiplying both sides of the equation by $e^{-in\Delta x k_h}/N$, summing over $n$, and exploiting the representation of the Kronecker delta provided before,
we obtain
\begin{align}
\ddot Q_{jh}(t)+\omega_{h}^2(t) Q_{jh}(t)=0.
\label{eql}
\end{align}
To solve it, we employ the multiple scale analysis. Details about this strategy, as well as the solutions of this equation, are reported in the next section. 
Once the Josephson energy returns to its original value at time $t_\textrm{f}$, $E(t_\textrm{f}) = E(0) \equiv E_0$ and remains constant, the solution of the equation of motion simply becomes:
\begin{align}\label{solf}
 Q_{jh}=A_{jh}e^{i\omega_{0h} t}+B_{jh}e^{-i\omega_{0h} t},
 \end{align}
where the coefficients $A_{jh}$ and $B_{jh}$ are determined by the continuity condition of each $Q_{jh}$ at the end of the motion. The explicit form of the coefficients $A_{jh}$ and $B_{jh}$ is calculated for the two LHTLs in the next section.

Note that the $input$ and $output$ modes determine the quantum vacuum at different time, and their relation is described by Bogoliubov transformations in Eq.\eqref{bt}.
Substituting Eq.\eqref{solf} into Eq.\eqref{p>0}, we achieve the form of output Fourier mode expansion of the field with respect to the input modes. Recombining all terms properly and comparing with respect to Eq.\eqref{bt} we observe that
\begin{align}
 \alpha_{jh}=&\sqrt{2\omega_{0h}} B_{jh},\nonumber\\
 \beta_{jh}=&\sqrt{2\omega_{0h}}A_{jh}.
\end{align}
These formulas relate both Bogoliubov coefficients to the coefficients $A_{jh}$ and $B_{jh}$.

\section{Multiple scale analysis}\label{MSA}
In this section, we present the solution strategy to solve Eq.\eqref{eql} for the two LHTLs.
This formalism is presented in \cite{bender1999advanced}. As first, we define a new ``time scale" $\tau=\eta t$ and expand $Q_{jh}(t)$ with respect to $\eta$:
\begin{align}\label{msa}
Q_{jh}(t)=& Q_{jh}^{(0)}(t,\tau)+\eta Q_{jh}^{(1)}(t,\tau),\nonumber\\
\ddot Q_{jh}(t)=&\partial_t^2 Q_{jh}^{(0)}(t,\tau)+\eta \left[2\partial_{\tau t}^2 Q_{jh}^{(0)}(t,\tau)+\partial_t^2 Q_{jh}^{(1)}(t,\tau)\right]
\end{align}
where $\partial_t^2\equiv\partial^2/\partial t^2$ and $\partial_{\tau t}^2\equiv\partial^2/(\partial \tau \partial t)$.
We now substitute Eq.\eqref{msa} into Eq. \eqref{eql} and solve this equation at different orders in $\tau$. At the zeroth order in $\tau$ we simply achieve
\begin{align}\label{zeroth}
\ddot Q_{jh}^{(0)}+\omega_{h0}^2 Q_{jh}^{(0)}=0,
\end{align}
whose solution is 
\begin{align}\label{sol1}
 Q_{jh}^{(0)}(t,\tau)=A_{jh}(\tau)e^{i\omega_{h0} t}+B_{jh}(\tau)e^{-i\omega_{h0} t}.
 \end{align}
From the continuity conditions for $Q_{jh}(t)$ we have the following initial conditions for $A_{jh}(\tau)$ and $B_{jh}(\tau)$:
\begin{align}\label{ic}
A_{jh}(0)=&0, \nonumber\\
B_{jh}(0)=&\frac{1}{\sqrt{2\omega_{h0}}}\delta_{jh}.
\end{align}
We now focus on \textit{circuit 1}, describing a quantum magnetic flux field with massless excitations and dispersion relation given in Eq.\eqref{ds1} of the main text.
Considering only the first order in $\tau$, Eq.\eqref{eql} becomes:
\begin{align}\label{firsto}
2\partial_{\tau t} Q_{jh}^{(0)}+\partial_t^2 Q_{jh}^{(1)}+\omega_{h0}^2 Q_{jh}^{(1)}+4\omega_{h0}^2\sin(\Omega t)Q_{jh}^{(0)}=0,
\end{align}
Substituting the solution of the zeroth order, Eq.\eqref{sol1}, into Eq.\eqref{firsto} we obtain
\begin{align}
\partial_t^2 Q_{jh}^{(1)}+\omega_{h0}^2 Q_{jh}^{(1)}=&-2\partial_{\tau t} Q_{jh}^{(0)}-4\omega_{h0}^2\sin(\Omega t)Q_{jh}^{(0)}\nonumber\\
=&-2i\omega_{h0}\left(({\partial_\tau}A_{jh})e^{i\omega_{h0} t}-({\partial_\tau}B_{jh})e^{-i\omega_{h0} t}\right)+2i\omega_{h0}^2\left(e^{i\Omega t}-e^{-i\Omega t}\right)\left(A_{jh}e^{i\omega_{h0} t}+B_{jh}e^{-i\omega_{h0} t}\right)\nonumber\\
=&-2i\omega_{h0}e^{i\omega_{h0} t}\left( ({\partial_\tau}A_{jh})-\omega_{h0}B_{jh}e^{i(\Omega-2\omega_{h0}) t}\right)+2i\omega_{h0} e^{-i\omega_{h0} t}\left( ({\partial_\tau}B_{jh})-\omega_{h0}A_{jh}e^{-i(\Omega-2\omega_{h0}) t}\right).
\end{align}
We now seek a solution of such equation without \textit{secularities}. Secularities are all those terms proportional to $e^{\pm i\omega_{h0} t}$, namely all those term that are already solution of the homogeneous equation. To avoid such secularities, we need the coefficients of $e^{\pm i\omega_{h0} t}$ to vanish, thereby obtaining the following set of differential equations:
\begin{equation}
    \left\{ 
\begin{array}{ll} 
\frac{\partial A_{jh}}{\partial\tau}-\omega_{h0}\delta(\Omega-2\omega_{h0})B_{jh}=&0,\\
\frac{\partial B_{jh}}{\partial\tau}-\omega_{h0}\delta(\Omega-2\omega_{h0})A_{jh}=&0,
\end{array}\right.
\quad\text{or}
\quad
\left\{ 
\begin{array}{ll} 
\frac{\partial A_{jh}}{\partial\tau}=&\omega_{h0}\delta(\Omega-2\omega_{h0})B_{jh},\\
\frac{\partial B_{jh}}{\partial\tau}=&\omega_{h0}\delta(\Omega-2\omega_{h0})A_{jh}.
\end{array}\right.
\end{equation}
These sets of equations are solved by differentiating one, substituting it into the other one, and exploiting the initial conditions in Eq.\eqref{ic}. Finally we obtain:
\begin{align}
A_{jh}=&\frac{1}{\sqrt{2\omega_{h0}}}\sinh\left(\omega_{h0}\tau\right)\delta_{jh}\delta(\Omega-2\omega_{h0}),\nonumber\\
B_{jh}=&\frac{1}{\sqrt{2\omega_{h0}}}\cosh\left(\omega_{h0}\tau\right)\delta_{jh}\delta(\Omega-2\omega_{h0}).
\label{ABml}
\end{align}
We proceed similarly for \textit{circuit 2}, describing a quantum field with massive excitations. At the zeroth order in $\tau$, the equation for $Q_{jh}$ is identical to Eq.\eqref{zeroth}, where now $\omega_{h0}$ corresponds to Eq.\eqref{ds2} of the main text at $t=0$.
At the first order, the equation is:
\begin{align}\label{firsto2}
2\partial_{\tau t} Q_{jh}^{(0)}+\partial_t^2 Q_{jh}^{(1)}+\omega_{h0}^2 Q_{jh}^{(1)}+\frac{4E_0}{C}\sin(\Omega t)Q_{jh}^{(0)}=0.
\end{align}
Moving the zeroth order to the right side and replacing the solution of the zeroth order, Eq.\eqref{sol1}, into Eq.\eqref{firsto2} we obtain
{\small
\begin{align}
\partial_t^2 Q_{jh}^{(1)}+\omega_{h0}^2 Q_{jh}^{(1)}=&-2\partial_{\tau t} Q_{jh}^{(0)}-\frac{4\tilde E_0}{C}\sin(\Omega t)Q_{jh}^{(0)}\nonumber\\
=&-2i\omega_{h0}\left(({\partial_\tau}A_{jh})e^{i\omega_{h0} t}-({\partial_\tau}B_{jh})e^{-i\omega_{h0} t}\right)+\frac{2i\tilde E_0}{C}\left(e^{i\Omega t}-e^{-i\Omega t}\right)\left(A_{jh}e^{i\omega_{h0} t}+B_{jh}e^{-i\omega_{h0} t}\right)\nonumber\\
=&-2ie^{i\omega_{h0} t}\left(\omega_{h0} ({\partial_\tau}A_{jh})-\frac{\tilde E_0}{C}B_{jh}e^{i(\Omega-2\omega_{h0}) t}\right)+2i e^{-i\omega_{h0} t}\left(\omega_{h0} ({\partial_\tau}B_{jh})-\frac{\tilde E_0}{C}A_{jh}e^{-i(\Omega-2\omega_{h0}) t}\right).
\end{align}
}
As done above, we seek solution without secularities, and therefore we need to solve the pair of differential equations
\begin{align}
\frac{\partial A_{jh}}{\partial\tau}=&\frac{\tilde E_0}{C\omega_{h0}}\delta(\Omega-2\omega_{h0})B_{jh}, \\
\frac{\partial B_{jh}}{\partial\tau}=&\frac{\tilde E_0}{C\omega_{h0}}\delta(\Omega-2\omega_{h0})A_{jh},
\end{align}
whose solutions are
\begin{align}
A_{jh}=&\frac{1}{\sqrt{2\omega_{h0}}}\sinh\left(\frac{\tilde E_0\tau}{C\omega_{h0}}\right)\delta_{jh}\delta(\Omega-2\omega_{h0}),\nonumber\\
B_{jh}=&\frac{1}{\sqrt{2\omega_{h0}}}\cosh\left(\frac{\tilde E_0\tau}{C\omega_{h0}}\right)\delta_{jh}\delta(\Omega-2\omega_{h0}).
\label{ABmv}
\end{align}

A crucial consequence of the duality between LHTLs and RHTLs is that, by repeating the same procedure and solving the equation of motions for RHTLs, we exactly obtain Eqs.\eqref{ABml} and Eqs.\eqref{ABmv} in the massless and massive cases respectively (with $\upsilon_{0h}$ instead of $\omega_{0h}$).

\end{document}